\documentclass[galaxies,article,submit,moreauthors,10pt,a4paper]{mdpi_no_lineno}  
\firstpage{1} 
\articlenumber{x}
\doinum{10.3390/------}
\pubvolume{xx}
\pubyear{2016}
\copyrightyear{2016}
\externaleditor{Academic Editor: name}
\history{Received: date; Accepted: date; Published: date}

 \theoremstyle{mdpi}
 \newcounter{thm}
 \newcounter{ex}
 \newcounter{re}
 \theoremstyle{mdpidefinition}

\Title{
Resolving the base of the relativistic jet in M87 at $6R_{\rm sch}$ resolution with global mm-VLBI
}

\Author{
Jae-Young Kim $^{1 *}$, 
Ru-Sen Lu $^{1}$,  
Thomas P. Krichbaum $^{1}$, 
Michael Bremer $^{2}$, 
J. Anton Zensus $^{1}$, 
R. Craig Walker $^{3}$, 
and the M87 collaboration}

\AuthorNames{Jae-Young Kim, 
Ru-sen Lu, 
Thomas P. Krichbaum, 
Michael Bremer, 
J. Anton Zensus, 
R. Craig Walker,
and the M87 collaboration}

\address{%
$^{1}$ \quad Max Planck Institut f\"{u}r Radioastronomie, Auf dem H\"{u}gel 69, D-53121 Bonn, Germany \\
$^{2}$ \quad Institut de Radio Astronomie Millim\'etrique, 300 rue de la Piscine, 
Domaine Universitaire, 38406 Saint Martin d' H\'eres, France \\
$^{3}$ \quad National Radio Astronomy Observatory, P. O. Box O, Socorro, NM 87801, USA  
}

\corres{Correspondence: jykim@mpifr-bonn.mpg.de; Tel.: +49-228-525-366}

\abstract{
M87 is one of the nearest radio galaxies with a central SMBH and a prominent relativistic jet.
Due to its close distance to the observer and the large SMBH mass, the source is one of the best laboratories to obtain strong observational constraints on the theoretical models for the formation and evolution of the AGN jets.
In this article, we present preliminary results from our ongoing observational study about the innermost jet of M87 at an ultra-high resolution of $\sim50\mu$as achieved by the Global Millimeter-VLBI Array (GMVA).
The data obtained between 2004 and 2015 clearly show limb-brightened jets at extreme resolution and sensitivity.
Our preliminary analysis reveals that the innermost jet expands in an edge-brightened cone structure (parabolic shape) but with the jet expansion profile slightly different from the outer regions of the jet.
Brightness temperatures of the VLBI core obtained from cm- to mm-wavelengths show a systematic evolution, which can be interpreted as the evolution as a function of distance from the BH.
We also adopt an alternative imaging algorithm, BSMEM, to test reliable imaging at higher angular resolution than provided by the standard CLEAN method (i.e. super-resolution).
A demonstration with a VLBA 7mm example data set shows consistent results with a near-in-time 3mm VLBI image.
Application of the method to the 2009 GMVA data yields an image with remarkable fine-scale structures that have been never imaged before.
We present a brief interpretation of the complexity in the structure.
}


\begin{document}

\section{Introduction}

The giant elliptical galaxy M87 is one of the nearest radio galaxies with a distance $d$ of only $16.7 \pm 0.5$ Mpc \cite{dist} and a large Super-Massive Black Hole (SMBH) mass $M_{\rm BH}$ of $(3-6.6) \times10^{9}M_{\odot}$ (\cite{mass},\cite{mass2}), 
although the exact mass is still controversial.
Owing to its proximity and the large BH mass, 1 milli-arcsecond (mas) on the sky plane corresponds only to $128R_{\rm sch}$ when we adopt $M_{\rm BH}=6.6\times10^{9}M_{\odot}$ (\cite{mass}), where $R_{\rm sch}$ is the Schwarzschild radius.
This is the best spatial resolution achievable for any extragalactic jet-hosting system.
Furthermore the M87 jet is transversely resolved and exhibits a limb-brightened structure, which contains a lot of information about the physical conditions (e.g. \cite{asada12_coll}, \cite{hsa_3mm}).
The close distance, the high black hole mass, and the uniquely limb-brightened structure therefore make M87 one of the best sources to answer fundamental questions about the formation and evolution of relativistic outflows in BH-accretion systems.

Accordingly, observing the inner-most region of the M87 jet with mm-VLBI is crucial to provide a connection between theoretical models and actual observations.
Very Long Baseline Interferometry (VLBI) has been so far the only way that allows us to zoom in close enough to the central engine. 
Especially, the Global Millimeter-VLBI Array (GMVA) at 3mm resolves the innermost jet at a $50 \mu$as scale, which corresponds to a spatial scale of only $\sim 6R_{\rm sch}$.
This resolution allows us to probe the geometry of the outflow near the jet launching region.
Furthermore the 3mm global VLBI observations also complement the ongoing 1mm VLBI experiments (with Event Horizon Telescope; EHT), that have even higher resolving power and aim to resolve event-horizon-scale structures, but still lack high-fidelity imaging capability (e.g. \cite{eht_sci}).

In this article we present preliminary results from our ongoing 3mm-VLBI observations of M87 with the GMVA. 
A general description of our data is given in Section 2. 
In Section 3 we describe the data analysis and possible interpretations.
Then we give the conclusions in Section 4.

\section{The 3mm VLBI observations of M87}

For this study we analyze a subset of a larger 3mm GMVA data set on M87. 
The data have been obtained since 2004 including a new high sensitivity data set from the latest GMVA observations performed in May 2015.
The 2004 and 2009 data had been already presented in \cite{krichbaum06} and \cite{krichbaum14} and were re-imaged for this study.
We also included one archival 3mm VLBI data set from VLBA+GBT observations in 2014 (BH 186; \cite{hsa_3mm}) in order to improve the time-sampling.
Table \ref{tab:data} summarizes basic information of the observations and the angular resolutions.
All the observations were performed in full-track -- the typical duration of the observations for the GMVA was $\sim$15 hours including the uptime of both European and US stations while it was roughly 8.5 hours for the VLBA+GBT observation (see \cite{hsa_3mm} for more details).
In 2014 and 2015, the data were recorded in dual polarization, while the earlier epochs observed in LCP only.
The GMVA data were correlated at Bonn by the MPIfR correlator.
The data were then fringe fitted and calibrated in the standard manners in AIPS. 
After this the data were imaged in Difmap using multiple steps of iterative phase and amplitude self-calibrations, until the rms noise level in the residual maps reached a minimum.
Fig. \ref{fig:allmaps} shows the preliminary jet images.
We also made a stacked image by convolving the individual images with a common restoring beam of $0.30 \times 0.11$ mas at the position angle of $0^{\circ}$, which is basically the same as the beam of the 2014 data, 
in order to display a persistent jet structure elongated in the E-W direction. 
The inner jet of M87 at $\leq1$ mas is obviously limb-brightened in all individual epochs, showing a complex expanding structure.

\begin{table}[H]
\caption{Summary of the observations and the data. 
Station abbreviations are: 
EB - Effelsberg, 
ON - Onsala, 
PV - Pico Veleta, 
PB - the phased Plateau de Bure interferometer,
GBT - the Green Bank Telescope, and 
VLBA - the 3mm VLBA stations (i.e. without Hancock and Saint Croix).
The restoring beam sizes are from the natural weighting.
(a) Brewster was not available.
(b) The observations in 2009 were conducted in full-track mode over two consecutive days.
(c) PB is limited to a bandwidth of 256 MHz due to 1Gbps of limited maximum data recording rate.
}
\small 
\centering
\begin{tabular}{cccc}
\toprule
\textbf{Date [yyyy/mm/dd]} & \textbf{Stations} & \textbf{Beam Size (mas, deg)} & \textbf{Bandwidth [MHz]} \\
\midrule
2004/04/19 & EB, ON, PV, VLBA (-BR$^{a}$) & $0.077 \times 0.310$, -11.69 & 128 \\
2005/10/15 & EB, ON, PV, VLBA & $0.071 \times 0.241$, -7.44 & 128 \\
2009/05/09--10$^{b}$ & EB, ON, PB, VLBA & $0.079 \times 0.274$, -12.22 & 128 \\
2014/02/26 & GBT, VLBA & $0.115 \times 0.305$, -10.5 & 512 \\ 
2015/05/16 & EB, ON, PV, PB, GBT, VLBA & $0.061 \times 0.275$, -8.87 & 512$^{c}$ \\ 
\bottomrule
\label{tab:data}
\end{tabular}
\end{table}

\section{Analysis and Discussion}

\subsection{The jet collimation at $< 100R_{sch}$ from the long-term time averaged stacked images}\label{ss:coll}

Our highest resolution and high sensitivity VLBI images ever obtained for the M87 jet allow us to reliably study how the innermost jet at $<100R_{\rm sch}$ is collimated. 
Here we show the analysis of evolution of the jet diameter versus the projected distance from the core using a stacked image under the assumption that the outer jet structure does not significantly change over time.
In order to gain sensitivity for the image analysis we convolved the VLBI images for 5 epochs (2004 to 2015) with a common circular beam of 0.1mas and then stacked the images.
The resulting image is shown in Fig. \ref{fig:stack_sr}.
In the analysis the map is rotated clockwise by $18^{\circ}$ so that one can easily slice the jet cross-section and fit two Gaussians to the two intensity peaks.
The slice calculation starts from $\sim$ 1 mas ($\sim$ 128$R_{\rm sch}$) from the core and makes slices at every pixel by moving down to 0.1 mas from the core which is the restoring beam size of the stacked image.
We then calculate the jet width $W$ at a given projected distance $z$ from the core from the separation of the two peaks of the 2 fitted Gaussians.
Similarly, we also calculate the apparent opening angle of the jet at each distance.
We fit a simple model of the jet diameter $W(z) \propto z^{k}$ to the data in order to quantitatively describe the transverse jet expansion.

\begin{figure}[H]
\centering
\includegraphics[width=12cm]{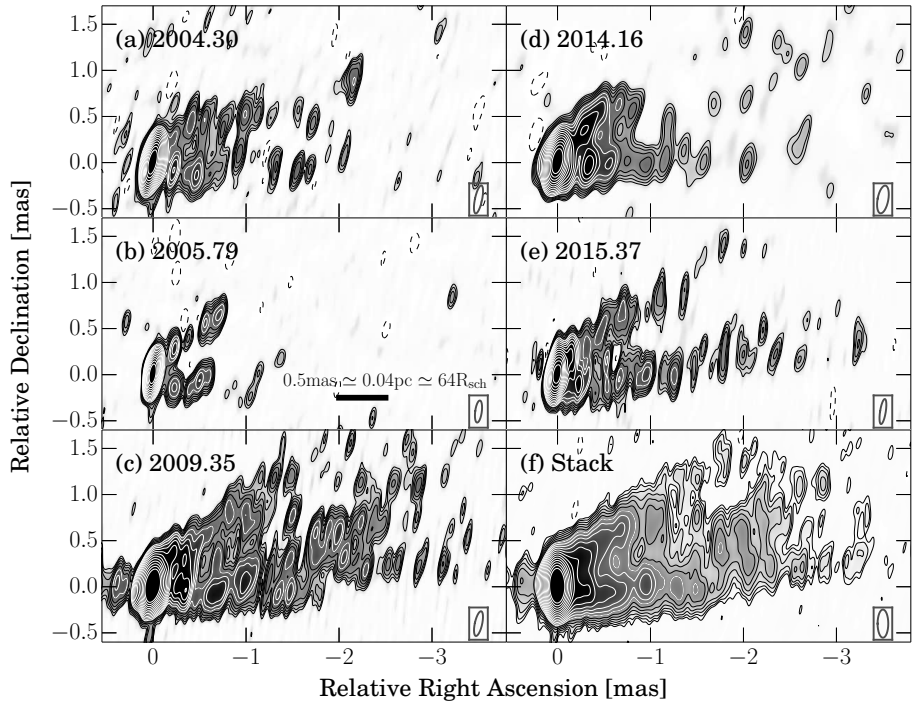}
\caption{
The preliminary 3mm VLBI images of M87 ordered in time. 
The beam is shown at the bottom right corner and the observing date is shown at the top left corner in decimal year format.
The contours and gray color show the total intensities.
Except panel (f), the lowest contour level corresponds to 1 mJy/beam and the contours increase by a factor of $\sqrt{2}$.
For panel (f) the lowest level is 0.4 mJy/beam. 
}
\label{fig:allmaps}
\end{figure}   

The result is shown at the left panel of Fig. \ref{fig:collimation_tb}.
Owing to the long time baseline of our data, the stacked image reveals a well-defined shape for the extended jet boundary.
Accordingly, our jet diameter measurements successfully reconstruct the inner jet collimation profile at high fidelity. 
The overall jet diameter at this scale is comparable to a nearly single-epoch result reported in recent work (see Fig. 9 of \cite{hsa_3mm}).
However, one significant difference between the result of this work and that of \cite{hsa_3mm} is that 
we do not see the strongly over-collimated shape of the jet base at $z=0.2-0.3$ mas.
This suggests that the diameter of the jet base could locally vary with time.
One of simplest explanations for the difference would be that an advection-dominated hot accretion flow (ADAF) near the BH is dynamically variable on time scales of at least several years
if the shrinking feature was caused by over-pressure from an ADAF as suggested by \cite{hsa_3mm}.
Furthermore, the derived power-law index $k = 0.52\pm0.01$ slightly differs from what different studies have obtained by analyzing larger-scale part of the same jet ($0.58\pm0.02$ \cite{asada12_coll}, $0.56\pm0.03$ \cite{hada13_coll}, $0.60\pm0.02$ \cite{mertens_thesis}).
These issues will be investigated in more detail in the future.

\begin{figure}[H]
\centering
\includegraphics[width=11.5cm]{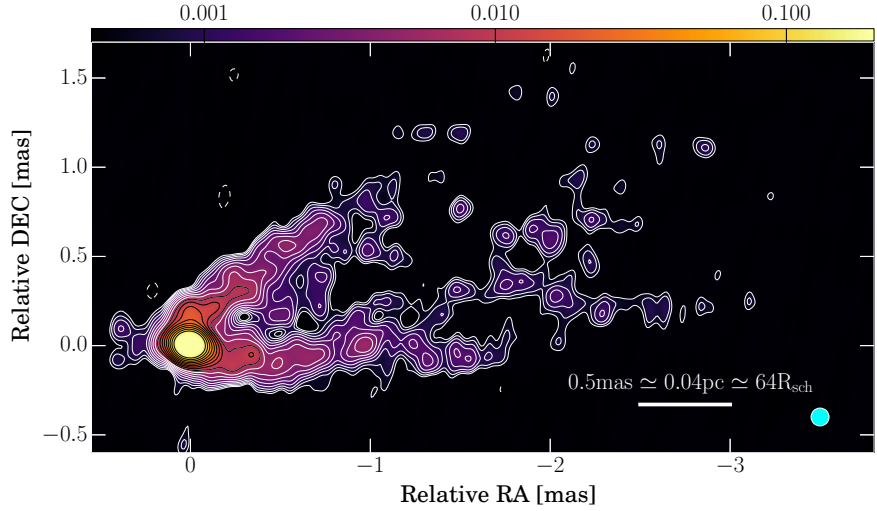}
\caption{
High-resolution image of M87 jet at 3mm obtained by stacking the five VLBI maps convolved with a common circular beam of $0.1 \times 0.1$ mas shown as the cyan-colored circle at the right bottom corner.
The colorbar indicates Stokes I intensity in units of Jy/beam.
The peak intensity is 522 mJy/beam and the off-source rms noise level is 0.12 mJy/beam, resulting in a dynamic range of around 4300.
Lowest contour level is 0.5 mJy/beam and the contour lines increase by a factor of $\sqrt{2}$.
}
\label{fig:stack_sr}
\end{figure}   

\begin{figure}[H]
\centering
\includegraphics[height=6.cm]{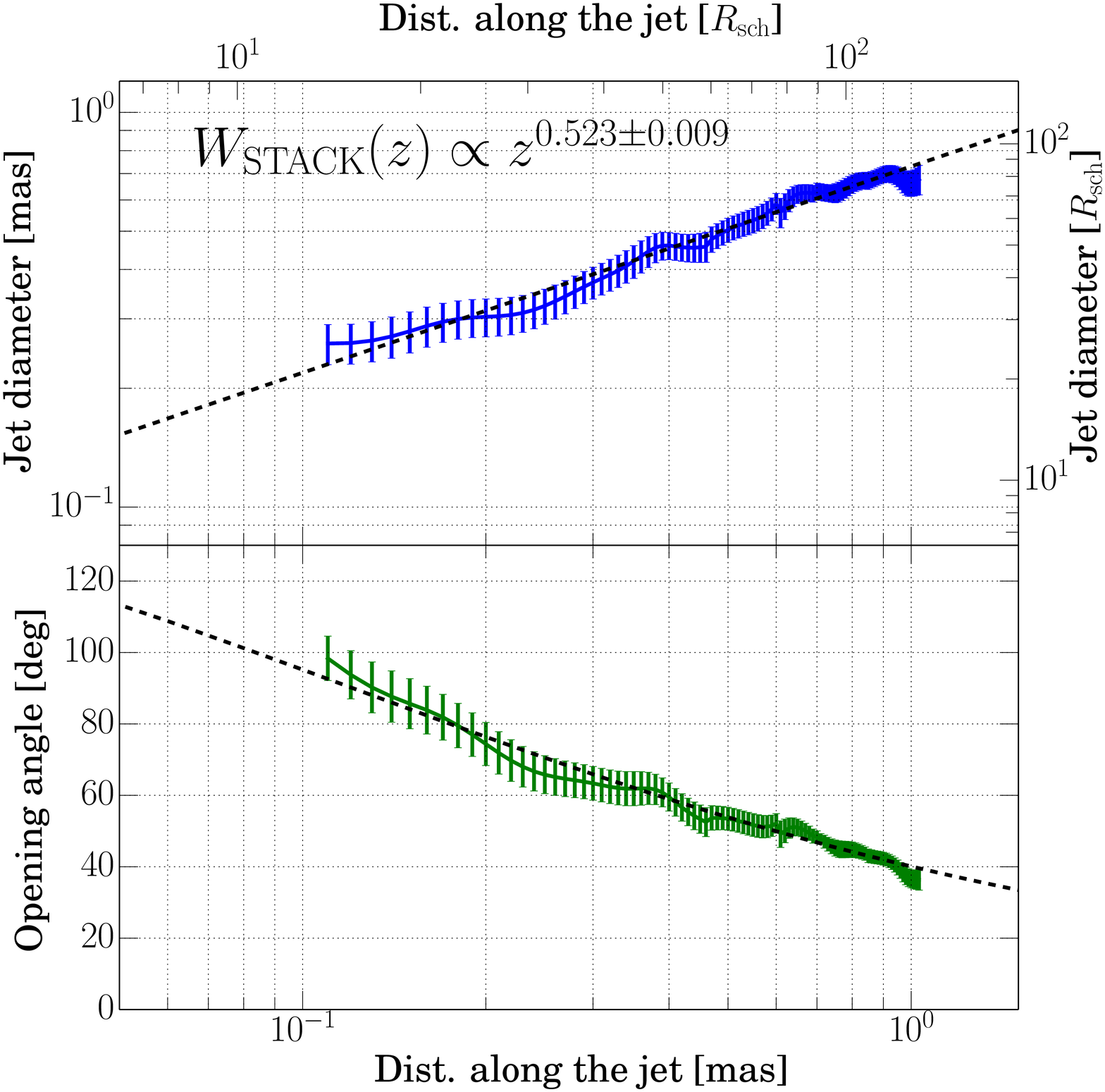}
\includegraphics[height=6.cm]{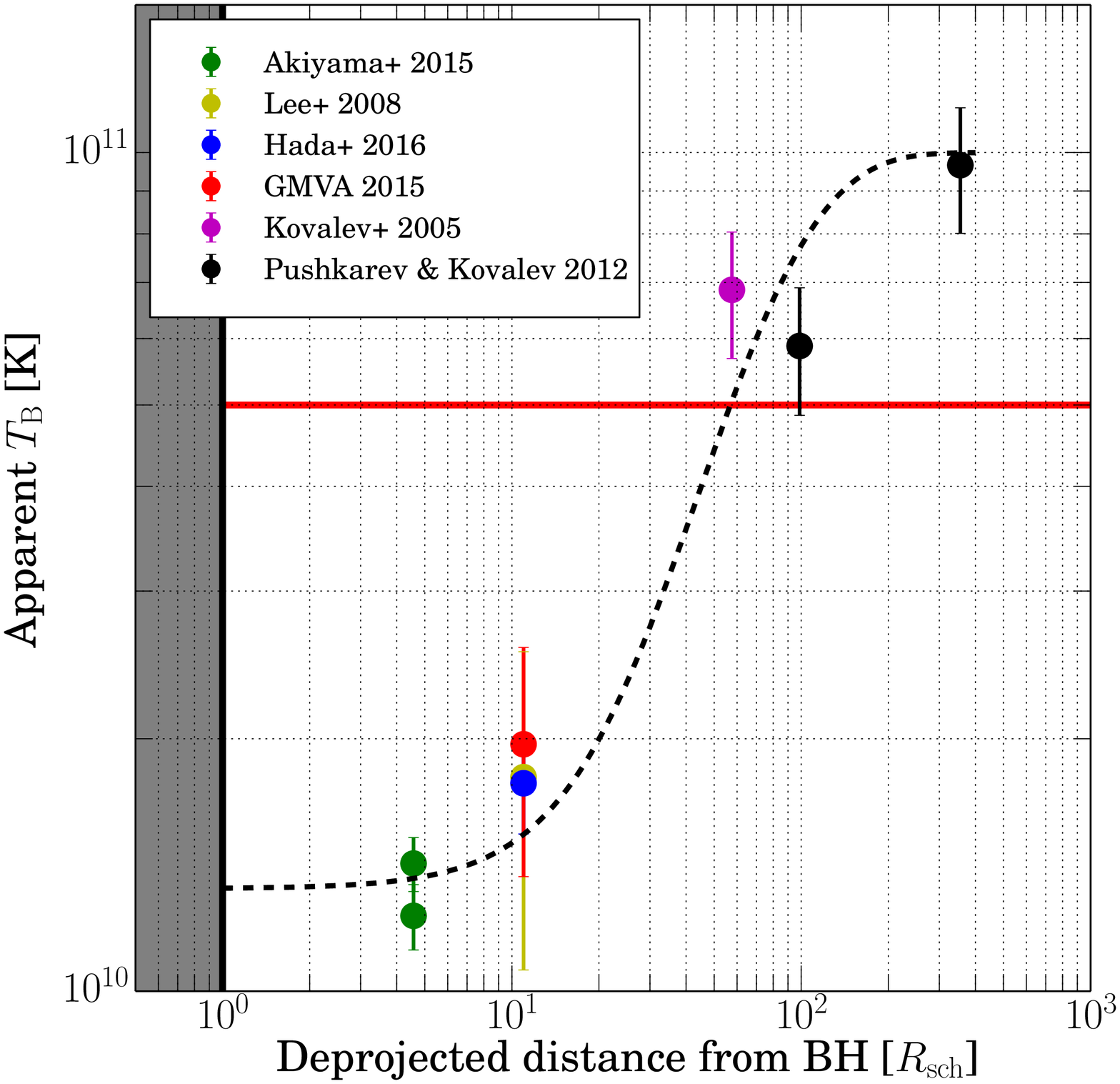}
\caption{
Left : the M87 inner jet collimation profile
with the dashed gray lines indicating the model fitting to the data.
Right : the frequency dependent brightness temperature evolution of the VLBI core displayed as function of the distance from the BH.
The gray-painted region indicates $1 R_{\rm sch}$, the red solid line the equipartition brightness temperature value of $T_{\rm B}=5\times10^{10}K$, 
the symbols for the origin of the data, including new GMVA data and data from literature
(\cite{akiyama_eht},\cite{hsa_3mm},\cite{kovalev05},\cite{tb_survey_lee},\cite{pushkarev12}).
The dashed line sketches a jet acceleration model.
}
\label{fig:collimation_tb}
\end{figure}

\subsection{Frequency-dependent evolution of $T_{B}$ in the M87 core}\label{ss:tb}
 
A recent study on the brightness temperature $T_{\rm B}$ evolution over several radio frequencies showed that the VLBI cores in compact AGN jets have systematically lower $T_{\rm B}$ at 3mm than at longer wavelengths \cite{tb_lee}.
We investigate this issue in particular for M87 in more detail.
For the VLBI core, we compiled a set of $T_{\rm B}$ measurements between 2.3, 8.6, 15.4, 86, and 230 GHz using available literature data and including our latest GMVA data set of 2015.
In a longitudinally pressure-stratified jet model, it is possible to obtain the absolute position of a VLBI core from the BH at a specific observing frequency (e.g. \cite{lobanov_coreshift}).
Therefore we can build a model describing the brightness temperature evolution as a function of the distance.
To simplify the modeling we adopt the frequency-distance relationship presented in \cite{hada11_coreshift} instead of assuming energy equipartition at the VLBI core.

\begin{figure}[H]
\centering
\includegraphics[width=11.5cm]{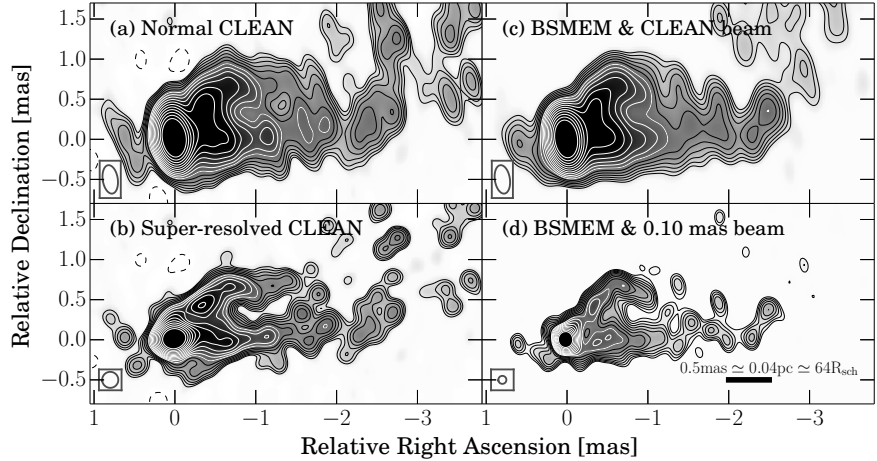}
\caption{
Demonstration of the imaging capability of BSMEM using real 7mm VLBA data.
(a) The image obtained from normal CLEAN method (beam : $0.19\times0.36$ mas, $8.59^{\circ}$).
(b) Super-resolved CLEAN image with a circular beam with FWHM of 0.20 mas.
(c) The source image reconstructed by BSMEM and convolved with the beam of the normal CLEAN result.
(d) A BSMEM map convolved with a circular beam of 0.10 mas in size.
The beam sizes are shown at the left bottom corners.
Except for panel (d), the minimum contour level is 1.0 mJy/beam. For panel (d), it is 0.5 mJy/beam.
The other parameters for the display are the same.
}
\label{fig:myMEM}
\end{figure}

The right panel of Fig. \ref{fig:collimation_tb} shows the variation of $T_{\rm B}$ versus deprojected distances of the core positions from the central BH with our preliminary model based on Eq. 10 of \cite{tb_lee}, which successfully describes similar brightness temperature variation of larger sample.
Apparently the observed $T_{\rm B}$ values are divided into two regimes -- lower ($\leq2\times10^{10}K$) and higher $T_{\rm B}$ ($\geq5\times10^{10}K$) for the inner and outer jet regions, respectively.
We consider two possible implications of this finding.
(1) Given the small distance scales from the BH probed by this analysis, it is possible to assume that the intrinsic brightness temperatures may remain the same over distances.
If this is true, the observed large apparent $T_{\rm B}$ in the outer regions could have been just Doppler-boosted by the Doppler factor $\delta$ which increases with larger jet speed.
This could be the case if the plasma flow in the inner region ($\leq100R_{\rm sch}$) is under strong acceleration.
In fact, such an acceleration is directly observed from the proper motion of bright emission features in the VLBA 43 GHz ''movie'' of the M87 jet on relatively larger distance scales
(\cite{walker_proceeding}, \cite{mertens_thesis}, \cite{mertens_paper}, \cite{walker_inprep}).
(2) In the other scenario, if the rapid inner jet acceleration is not taking place within $\leq100R_{\rm sch}$, the observed apparent $T_{\rm B}$ should have not been strongly Doppler-boosted and the intrinsic $T_{\rm B}$ at cm-wavelength should be close to the observed values. 
If this is the case, the transition of the lower $T_{\rm B}<2\times10^{10}K$ at mm-regime to the higher $T_{\rm B}$ of $>5\times10^{10}K$ at cm-wavelengths suggests that the dominant energy budget of the inner M87 jet is changing as the jet plasma propagates outward. 
In other words, the initially magnetically-dominated inner jet is heating the internal plasma by rapidly dissipating the magnetic energy.

However it should be noted that there are several uncertainties involved in this simple modeling, especially regarding how to convert the core position at a given wavelength to the distance when 
the observing radio frequencies are higher than 43 GHz.
For example, the boundary of the innermost jet of M87 is not conical as shown in Subsection \ref{ss:coll} and the VLBI core at $\nu_{\rm obs}>43$ GHz is probably magnetic-energy dominated as discussed in \cite{kino_14} and \cite{kino_15}.
In such a case, it is more difficult to determine the exact locations of the mm-VLBI core with respect to the BH by extrapolating the relationship obtained from lower frequencies.
We are currently working on improving the modeling.

\begin{figure}[H]
\centering
\includegraphics[width=12cm]{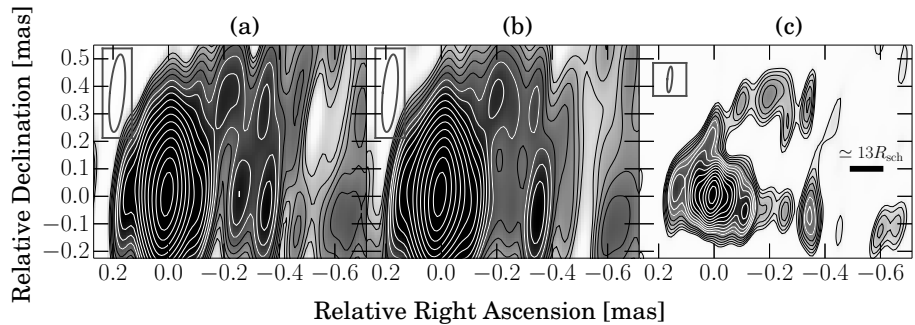}
\caption{
CLEAN and BSMEM imaging results for the 2009 May GMVA visibility data.
For each panel,
(a) normal CLEAN (beam : $0.049\times0.290$ mas, $-5.99^{\circ}$), 
(b) BSMEM image with the same beam as (a), and
(c) super-resolved BSMEM image (beam : 1/3 of the CLEAN beam size).
In panel (a) and (b) the lowest contour level is 2.0 mJy/beam, which roughly corresponds to 7 sigma for the CLEAN map.
For panel (c) the lowest level is 1.0 mJy/beam.
}
\label{fig:myMEM2}
\end{figure}

\subsection{Fine-scale structure around the nuclear region}\label{ss:mem}

The angular resolution $\theta$ of a VLBI experiment is proportional to the observing wavelength $\lambda$ divided by the maximum baseline length $D$; $\theta \approx \lambda/D$.
In more detail, however, the angular resolution is not uniquely fixed for exactly similar data.
For example, it changes with different signal-to-noise (e.g. \cite{lobanov_limit}), different visibility weighting (e.g. \cite{gomez_RA}), and different imaging methods (e.g. \cite{lu_imaging}).
Recent numerical tests on observed and simulated data (e.g. \cite{chael_bsmem}) have shown that alternative imaging methods such as the Maximum Entropy Method (MEM) are able to achieve even higher angular resolution than the standard CLEAN, providing ways for the VLBI experiments to obtain better resolutions than previously considered.
Being motivated by the possibility of detecting further fine-scale structures from our 3mm VLBI data, we adopt the Bi-Spectrum Maximum Entropy Method (BSMEM; \cite{bsmem_original}) for the super-resolution imaging.

We first demonstrate the imaging method using an archival data set from 7mm VLBA observations of M87 performed in 2014 March (BH 186; \cite{hsa_3mm}).
We created four images of the jet using the same visibility data to compare CLEAN and BSMEM. The imaging results are shown in Fig. \ref{fig:myMEM}.
The CLEAN image displays a limb-brightened jet with a weak counter-jet feature.
But, due to the relatively large beam size especially in the N-S direction, it is difficult to see whether the inner region of the flow ($\leq0.4$ mas) is also limb-brightened.
The super-resolved CLEAN map provides a hint for a complex geometry near the core.
When it comes to the details such as the jet opening angle, however, the interpretation of the structure is not unambiguous, especially when compared with a jet structure obtained by 3mm VLBA+GBT observations performed roughly one month earlier 
(see Fig. 4 of \cite{hsa_3mm}).
On the other hand, BSMEM reconstructed much more reliable structure at much higher resolution.
To give a comparison, we first convolve the BSMEM source model image with the same CLEAN beam as panel (a).
As panel (a) and (c) show, the two images obtained from different methods agree well. 
When the convolving beam for the BSMEM model is 0.10 mas (panel d), the jet clearly shows limb-brightening below 0.4 mas with a hint for a wide opening angle of the counter jet.
Actually, the inner jet structure is quite a bit closer to the 3mm VLBA+GBT image (Fig. 4 of \cite{hsa_3mm}) than the super-resolved CLEAN map.
Therefore we conclude that BSMEM can achieve significantly better angular resolution than CLEAN with super-resolution.

Following the above demonstration, we directly applied BSMEM to the 2009 GMVA data and we show the result in Fig. \ref{fig:myMEM2} by zooming into the central region.
We find that BSMEM again reconstructs an image consistent with the CLEAN result when the original CLEAN beam is used (panel a $\&$ b).
The super-resolved BSMEM image however presents complex filamentary system consisting of (a) an ultra-compact intensity peak which is located in the extended southern rail and (b) arc-shaped northern rail (panel c).
Based on the previous demonstration, we believe that the obtained fine-scale structure is real.
To our knowledge, this is the highest fidelity image of the M87 jet successfully obtained at this angular resolution.

The new image has several important scientific implications. For example, it can be seen easily that the bright nuclear emission region highlighted in white contours extends much more in the E-W direction than in the N-S direction (panel c).
This implies that the northern and southern rails are highly asymmetric in terms of their brightness.
A simple explanation is that more material was being injected toward the southern side when the jet was observed and a ``component'' appeared as a result. 
Perhaps the southern rail is more strongly Doppler-boosted because of different speeds between the two limbs or the jet rotation (e.g. \cite{mertens_paper}) under the assumption that the two rails are intrinsically symmetric.
The stacked 3mm VLBI image in Fig. \ref{fig:allmaps} shows that the southern rail could be slightly brighter in general.
However, temporal changes in the brightness ratio between the two rails seen in Fig. \ref{fig:allmaps} and also in the 7mm VLBA images (\cite{walker_inprep}) make the assumption of symmetry to be difficult to be justified.

We also point out that the bright southern rail and the compact core size complicate the determination of the exact position of the jet apex or the BH, at least for this epoch.
As mentioned above, the basic assumption is that the intensity peak corresponds to the true VLBI core and this idea works successfully on larger scales (e.g. \cite{hada11_coreshift}).
Nonetheless, on the spatial scales that we are probing now, the bright extended emission could blend the core with the jet, forming the intensity peak offset from the central engine.
Also, if the intensity peak is the location of the central engine, the position angle evolution of the northern rail would be difficult to understand.
Specifically, the base of the northern rail should have been launched from the core at a position angle of $\sim 15^{\circ}$ with respect to north to explain the observed jet base geometry. 
But the resulting apparent jet opening would be uncomfortably large. 
Another possible jet apex position in this perspective is the point of intersection of the two rails, 
which is $<0.1$ mas eastward of the intensity peak since this location at least explains better the observed jet morphology, 
i.e. the smooth and continuous, semi-parabolic shape of the expanding jet envelope.

\section{Conclusions}

In this article we have presented our ongoing study on the innermost jet of M87 with mm-VLBI and a demonstration of super-resolution imaging using the BSMEM algorithm.
Preliminary results from the analysis of our 3mm VLBI data showed 
a well defined, transversely expanding jet profile, 
significant evolution in the brightness temperature,
and highest resolution image of the jet base that has ever been made.
We are now analyzing details of the individual epochs along with long-term structural changes in an effort to better constrain the physical conditions.
Further super-resolution imaging of the rest of the 3mm VLBI data with BSMEM is also in progress to help this goal.

\vspace{6pt} 

\acknowledgments{
J. -Y. Kim is supported for this research by 
the International Max-Planck Research School (IMPRS) for Astronomy and Astrophysics
at the University of Bonn and Cologne.
This research has made use of data obtained with the Global Millimeter VLBI Array (GMVA), 
which consists of telescopes operated by the MPIfR, IRAM, Onsala, Metsahovi, Yebes, and the NRAO. 
The data were correlated at the correlator of the MPIfR in Bonn, Germany. 
The Very Long Baseline Array and the Green Bank Telescope are instruments of the National Radio Astronomy Observatory, 
a facility of the National Science Foundation operated under cooperative agreement by Associated Universities, Inc.
}

\bibliographystyle{mdpi}

\renewcommand\bibname{References}

\end{document}